# CARLA: A Convolution Accelerator with a Reconfigurable and Low-Energy Architecture

Mehdi Ahmadi, Shervin Vakili, and J.M. Pierre Langlois, *Member, IEEE*

*Abstract*— Convolutional Neural Networks (CNNs) have proven to be extremely accurate for image recognition, even outperforming human recognition capability. When deployed on battery–powered mobile devices, efficient computer architectures are required to enable fast and energy-efficient computation of costly convolution operations. Despite recent advances in hardware accelerator design for CNNs, two major problems have not yet been addressed effectively, particularly when the convolution layers have highly diverse structures: (1) minimizing energy-hungry off-chip DRAM data movements; (2) maximizing the utilization factor of processing resources to perform convolutions. This work thus proposes an energy-efficient architecture equipped with several optimized dataflows to support the structural diversity of modern CNNs. The proposed approach is evaluated by implementing convolutional layers of VGG-Net-16 and ResNet-50. Results show that the architecture achieves a Processing Element (PE) utilization factor of 98% for the majority of 3×3 and 1×1 convolutional layers, while limiting latency to 396.9 ms and 92.7 ms when performing convolutional layers of VGGNet-16 and ResNet-50, respectively. In addition, the proposed architecture benefits from the structured sparsity in ResNet-50 to reduce the latency to 42.5 ms when half of the channels are pruned.

## I. Introduction

DEEP LEARNING approaches based on Convolutional Neural Networks (CNNs) have demonstrated superior results in several vision tasks including image classification [1]-[3], object detection [4]-[6], and activity recognition [7], [8]. CNNs are constructed by stacking multiple convolutional layers followed by some fully connected (FC) layers and a classifier. Deep CNNs with numerous convolutional layers have proven to achieve high classification accuracy [3]. Consequently, the number of convolutional layers in modern deep CNNs has increased dramatically in recent years. For example, between AlexNet [1], the winner of ImageNet Large Scale Visual Recognition Challenge (ILSVRC) [10] in 2012, and ResNet [3], the winner in 2015, the number of convolutional layers increased by a factor of more than 20×. This increase allowed ResNet to outperform human level accuracy at the expense of a substantial growth in the number of parameters and multiply-accumulation (MAC) operations.

The massive data requirements and computational complexity pose significant deployment challenges of deep CNNs on energy-constrained portable devices. The required data, including network parameters, input image and intermediate results, is normally too large to fit within internal memories. Thus, using external memories such as DRAMs is considered inevitable. Since DRAM accesses consume orders of magnitude more energy than any other function [11], increasing local data reuse, which limit data transfers to and from DRAM, can significantly reduce energy consumption.

In addition, many vision tasks demand CNN inference times within hundreds of milliseconds per image [12], [13]. Acceleration can be achieved through computational parallelism. However, CNN layers have diverse convolution configurations such as different filter and feature map sizes. Thus, a major challenge in designing CNN accelerators is to effectively map these diverse convolutional layers into a fixed computing architecture.

Research on CNN complexity reduction and computational acceleration has evolved extensively during the past few years [15]-[31]. At the architecture level, a large number of domain-specific neural network processors have been introduced for accelerating deep learning tasks on servers and datacenters [16]-[19]. Strict energy and power consumptions constraints are not demanded in such non-mobile applications. For example, Google TPU [19] and DaDianNao[17] have power consumptions of 40 W and 16 W, respectively. Some previous works introduced FPGA-based CNN accelerators [30], [31] and these designs typically consume power in the order of tens of watts.

Another approach is to design accelerators that only consume a few hundreds of milliwatts to fit within the power budget of mobile devices [20]. This class of accelerators can be used by host mobile processors to perform the convolution computations. Several recent works have proposed energy efficient accelerators for state-of-the-art CNNs, on Application-Specific Integrated Circuit (ASIC) [21]-[29]. Tu et al. [23], proposed the Deep Neural Architecture (DNA) for the convolutional and fully connected layers of AlexNet. Moons et al. [24] presented Envision, a low-power scalable accelerator, which exploits Dynamic Voltage Accuracy Frequency Scaling (DVAFS) to achieve energy efficiency. The architecture configures multipliers into different word-lengths based on the accuracy requirements of convolutional layers in AlexNet and VGGNet. Chen et al. [25] proposed an architecture, called Eyeriss, to compute the convolutional layers of AlexNet and VGGNet. Eyeriss includes a 2D spatial array of PEs connected through a network-on-chip (NOC), and it uses a row-stationary dataflow and a large global buffer to store intermediate data to reduce the number of off-chip DRAM accesses. Ardakani et al. [26], proposed a FC-inspired dataflow (FID) customized for VGGNet-16 and

M. Ahmadi, S. Vakili, and J.M.P. Langlois are with the Department of Computer and Software Engineering, Polytechnique Montréal, Montréal, QC H3T 1J4, Canada (e-mail: mehdi.ahmadi@polymtl.ca; shervin.vakili@polymtl.ca; pierre.langlois@polymtl.ca).

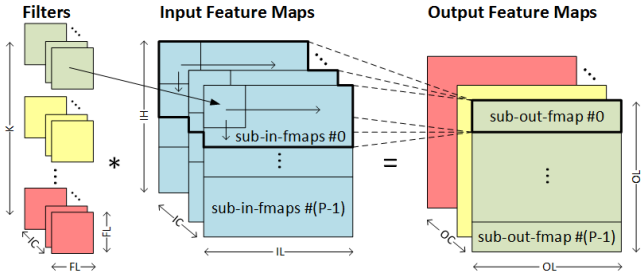

Fig.1 Convolution operation in a convolutional layer.

VGG-like networks based on computational cores of fully-connected layers. In that work, convolutional layer operations are treated as a special case of fully connected ones where a set of weights is shared among all neurons. The architecture utilizes computational units called Tiles, each consisting of sets of neurons and a weight generator unit that provides the Tiles with neuron weights using a shift register. ZASCA [28] is built upon FID, it supports more filter sizes and is able to compute the convolutional layers of ResNet-50.

Recently, Ahmadi et al. [29] introduced a convolution accelerator with a serial accumulation dataflow. That design only supports 3×3 convolutions of VGGNet-16. In this paper, we propose a new convolution accelerator with a reconfigurable and low-energy architecture, called CARLA, which supports efficient computation of various convolutional layers with different filter sizes. CARLA handles the computational diversity of different layers using several operating modes with distinct computational dataflows. The main contributions of this paper are as follows.

- The paper introduces CARLA, a convolution accelerator that integrates efficient dataflows/architectures for 1×1 and 3×3 convolutions with a low reconfiguration cost. In the 3×3 convolution mode, CARLA uses a serial accumulation dataflow by cascading PEs in each parallel unit while in 1×1 mode, PEs work independently. In addition, CARLA swaps the input feature and filter weight data movement when performing 1×1 convolution mode compared to the 3×3 mode. This enables the architecture to benefit from significant input data reuse in 1×1 convolution where the spatial size of filters shrinks to only a single weight in each channel.
- The paper proposes a new pipelining scheme to increase on-chip data reuse by taking advantage of feedbacks paths. In addition, this pipelining scheme provides the flexibility to switch data movement for different convolution modes.
- The paper proposes a new on-chip memory structure to support different convolutions. Moreover, it employs a mechanism that allows concurrency of computation with SRAM-DRAM data exchanges. To maximize memory efficiency, wider on-chip memories are employed to store the intermediate results while narrower on-chip memories are used to temporarily accommodate the final results.
- The paper provides performance analysis for each operation mode by illustrating the relation between the execution time, number of DRAM accesses, and the proportion of time that PEs actively contribute in computation, with the characteristics of the proposed architecture and CNN models.

- We evaluate CARLA on the convolutional layers of ResNet-50 which have very diverse specifications in the input data size (from 224×224 to 7×7), filter size (7×7, 3×3, and 1×1), number of filters (from 64 to 2048), input channels (64 to 2048) and filter strides (1 and 2). In addition, we show that CARLA is compatible with structured sparse CNN models and that it can efficiently take advantage of filter sparsity to reduce the computation time and number of memory accesses.

## II. DEFINITIONS AND BACKGROUND

### A. Convolutional Layers in Deep CNNs

Fig.1 shows the convolution operation in a convolutional layer. Each convolutional layer takes a 3-D input and a set of 3-D filters and generates a 3-D output. The 3-D input, also called input feature maps (in-fmaps), has size $IL \times IL \times IC$, where $IL$ is the length and $IC$ is the number of channels of the input. Each input element is called an input feature and each 2-D plane of size $IL \times IL$ is called a feature map. The input feature maps are convolved with 3-D filters of size $FL \times FL \times IC$ where $FL$ is the filter length. The dot product of the filter weights and the corresponding input features produces partial sums that are accumulated into output feature maps (out-fmaps). The 3-D out-fmaps are of size $OL \times OL \times OC$ where $OL$ is length of the output, while $OC$ is the number of output channels. Each output channel is the result of convolving the input with one filter. Therefore, with $K$ filters in a convolutional layer, the number of output channels is equal to the number of filters, i.e., $OC=K$. The convolution computation in a CNN layer is given by:

$$y_k(m,n) = b^k + \sum_{c=0}^{IC-1}\sum_{j=0}^{FL-1}\sum_{i=0}^{FL-1} x_c(mS+j, nS+i) \times w_c^k(j,i)$$

$$0 \leq m, n \leq OL, 0 \leq k \leq OC, OL = (IL - FL + 2Z)/S + 1. \quad (1)$$

where $x$, $w$, $y$ and $b$ are the elements in in-fmaps, filter, out-fmaps and bias matrices, respectively. $S$ denotes the filter stride and $Z$ is the number of zero pads [14].

### B. Architecture Design Challenges

When computing deep CNN inference in hardware, direct mapping of all computations into hardware resources of a single chip is not feasible due to the massive complexity of the associated computations. Existing designs commonly use a processing engine to perform a portion of computations at a time [23]. The engine is reutilized in a serial fashion to complete the entire computations.

Fig.1 shows the required computations in one convolutional layer using an example partitioning solution. In this example, each output channel is divided into $P$ sub-out-fmaps. Each sub-out-fmap is the result of convolving one filter with the associated portion of the input features, called sub-in-fmaps.

The large number of weights, input pixels and intermediate data cannot fit within internal memories and thus, utilizing external DRAMs is inevitable. Among all the required operations in hardware acceleration, the DRAM access is the most expen-sive one in terms of energy consumption. For instance, the relative energy consumption of a 32-bit DRAM access in 45 nm technology is 200× greater than a MAC operation [15], [33]. Hence, it is proportionally beneficial to maximize data reuse inside the chip to avoid DRAM accesses.

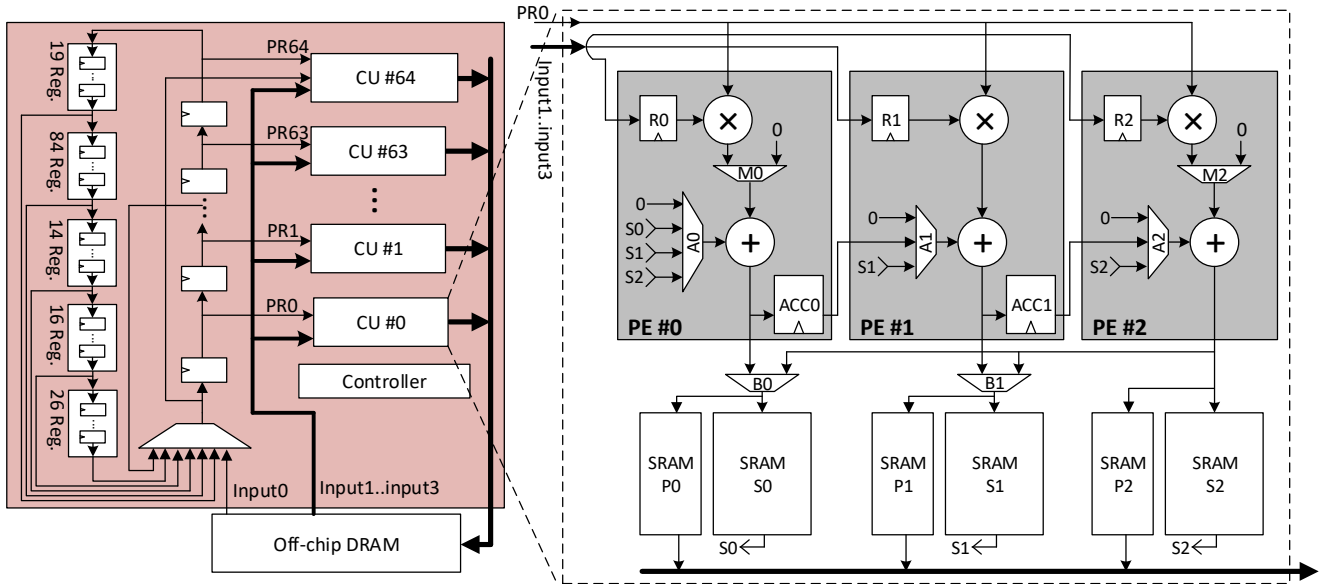

Fig. 2 The architecture of CARLA.

A straightforward solution is to utilize larger on-chip SRAMs to reduce the costly off-chip DRAM accesses at the expense of additional silicon area. In resource-limited ASIC designs, however, an optimized dataflow is necessary to minimize data movements between off-chip and on-chip memories while respecting resource constraints. When increasing the number of MACs, another challenge is to effectively map the computations onto the available MAC resources. Consequently, a significant issue in several existing designs is that they are not able to reach their advertised peak performance with a fixed architecture when executing modern CNNs [27].

## III. THE CARLA ARCHITECTURE

In a CNN inference accelerator, computation of each layer starts with fetching the weights and input pixels from the off-chip DRAM. The convolution operations are then performed, and the generated output features are stored back to the DRAM. These outputs are fed to the next layer as inputs.

CARLA is composed of a set of $U+1$ cascaded convolution units (CUs) and a controller. The controller handles the flow of the data transfers and assigns computations to the CUs. Each CU contains $N$ processing elements (PEs) to perform convolution operations, except the last one, CU #$U$, which can contain more PEs. Fig. 2 shows the proposed architecture configured for the case of ResNet models. Since the number of filters in ResNet models is divisible to 64 and 3×3 filters are more widely used compared to other large filter sizes, $U$ is set to 64 and $N$ is set to 3. In this case, CU #64 contains four PEs and all other CUs contain three PEs. CARLA has four input buses, called Input #0 to Input #3, coming from the external DRAM. Input #0 is buffered in a set of $U+1$ pipelined registers, each feeding one CU. Other buses, including Input #1 to Input #3, are connected to the inputs of all the CUs. The four input buses may carry weights or input features, interchangeably. As will be discussed later, this capability allows maximizing utilization of PEs for various operating modes.

The left side of Fig. 2 illustrates the pipeline structure designed for CARLA. Pipelining allows the sharing of input features among parallel units [26], [29]. The proposed pipeline has several feedback paths that allow the existing data inside the pipe to return into the pipeline input. In 3×3 convolution, some of the input feature rows are used in different computation steps. The feedback paths enable multiple reuse of those common data after the first fetch, instead of repeated re-fetching from the off-chip DRAM. This feedback mechanism is not used in 1×1 convolutions.

Fig. 2 also illustrates the internal structure of one CU when $N$=3. Inside a CU, each PE can perform one Multiply-and-Accumulate operation in each clock cycle. Each PE contains one internal register (e.g. R0 in PE #0) that provides the first operand of the multiply operation. These registers can be loaded with values from the CU inputs. The second multiplier operand is shared among all PEs of a given CU. This operand is provided from the pipelined registers.

Fig. 3 and Fig. 4 show the configuration of CARLA for 3×3 and 1×1 convolutions, respectively. Unused resources for each mode are shown with greyed-out lines. We point out to different parts of PE #0 by blue arrows.

The PEs can be connected in series to compute the convolution with 3×3 filter sizes (Fig. 3), or they can process independent data for 1×1 filters (Fig. 4). The result of the MAC operation is either stored in an accumulator (ACC0 and ACC1) when PEs are connected in series, or directly stored in SRAMs when the PEs process independent data.

Several multiplexers route the signals inside a PE and between the PEs of a CU based on the operating mode and the working state. The first and the last PE have special multiplexers, MUX M0 and MUX M2, which enable discarding and replacing the results of multipliers with zero. This feature facilitates handling of zero-pads at spatial borders of in-fmaps. Based on the selected dataflow (described in Section III.A.1), these stored values will be used later to compute convolution results. Three other multiplexers, named MUX A0, MUX A1, and MUX A2, provide flexibility to handle various convolution configurations required in different modes.

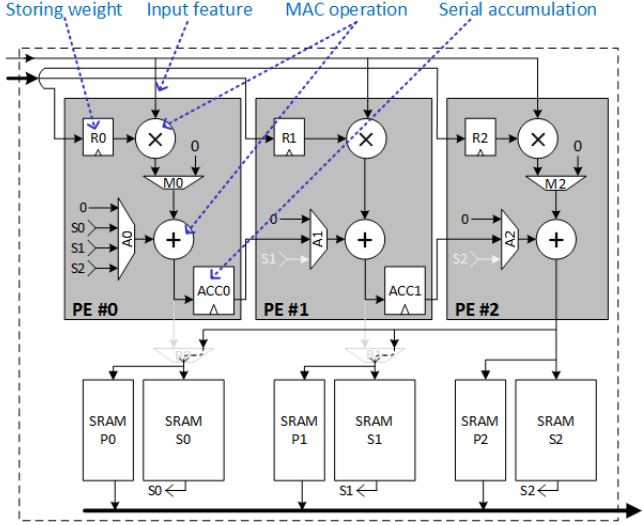
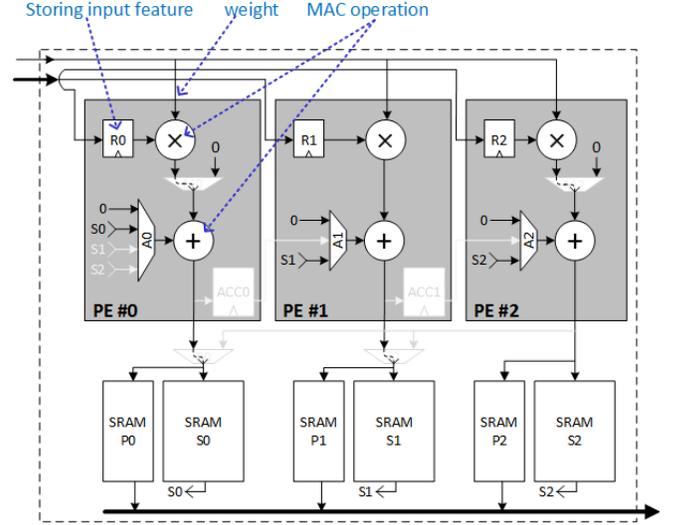

Fig. 3 The configuration of a CU for 3×3 convolution.

Fig. 4 The configuration of a CU for 1×1 convolution.

In the case of deep CNNs, on-chip SRAMs are generally not large enough to store entire out-fmaps. Thus, after computing a part of the out-fmaps, i.e., sub-out-fmaps, the architecture must free up the SRAMs by transferring the contents to the off-chip DRAM. On the other hand, the PEs need to write in the SRAMs every clock cycle. To enable the architecture to operate uninterruptedly, we propose to use a pair of SRAMs. While the PEs store partial results in one set of wide SRAMs, the narrower paired SRAMs receive the generated output features at the end of computation cycles. After completing each computation cycle, the contents of the narrower SRAMs are gradually transferred to the off-chip DRAM.

In CARLA, while PE #0 and PE #1 only have access to one pair of SRAMs, PE #2 can access all SRAMs. As will be discussed in Section III.B, each PE writes its produced partial sum directly into its allocated SRAMs when computing 1×1 convolution. However, in 3×3 convolution, the partial sums of the three PEs are added together and the result, called partial result, is obtained in PE #2 and stored in the SRAMs. Therefore, PE #2 has write access to all SRAM pairs to enable utilization of all SRAM resources in 3×3 convolution. The two multiplexers MUX B0 and MUX B1 handle the write access modes. The following subsections describe how CARLA performs the 1×1, 3×3, and 7×7 ResNet convolutions.

### A. 3×3 Convolution Mode

This section describes how CARLA performs 3×3 convolution efficiently. The steps to generate a single sub-out-fmap (see Fig.1) are described using the proposed configuration for the ResNet models, i.e., $U$=64 and $N$=3. For the 3×3 convolution case, CUs #0 to #63 each compute one output channel of the out-fmaps. Thus, in this section, we only explain the functionality of CU #0, while CU #1 to CU #63 operate in a similar way. CU #64 is not functional in 3×3 convolutions.

*1) Description of CU Functionality Using an Example:* This subsection describes the functionality of CU #0 using an example from the ResNet models, where a matrix of in-fmaps of size 56×56×64, with zero padding of 1, is convolved with 64 filters of size 3×3×64 with stride of 1. The out-fmaps matrix has a size of 56×56×64. The computation of the output channel $n$ is performed in CU #n. In this example, there are 56×56=3136 output features per output channel. In CARLA, each CU has a pair of SRAMs with 224 words. Thus, the out-fmap is divided into 3136/224=14 sub-out-fmaps of size 4×56. To produce 4 sub-out-fmaps, the sub-in-fmaps are divided into blocks of 6 rows. To summarize, CUs convolve sub-in-fmaps of size 6×56×64 with filters of size 3×3×64 to compute sub-out-fmaps of size 4×56. By repeating this process 14 times, all out-fmaps of one output channel are computed by the corresponding CU.

Fig. 5 shows the proposed dataflow to perform 3×3 convolution in CU #0. Due to zero padding of 1, the convolution of the first filter row with the first row of the sub-in-fmaps is not required. In clock cycle #0, the registers R0, R1 and R2 are loaded with weights of the first filter row. The following clock cycles are separated into three categories: routine computing cycles, in-fmap boundary handling, and row filter and input channel change.

**Routine computing cycles:** In clock cycle #1, the first input feature, $x_0(0,0)$, is fed to the input PR0. This input feature is multiplied to the filter weights of the first row, $w_0(0,0)$, $w_0(0,1)$ and $w_0(0,2)$, and the results are stored in the accumulators ACC0 and ACC1. The third adder computes $x_0(0,0) \times w_0(0,2)$, but this value does not contribute in any required computation. Therefore, it is discarded and replaced by zero using the multiplexer MUX M2.

In clock cycle #2, the second input feature, $x_0(0,1)$, appears in PR0, while the registers maintain their previous values. As a result, the third adder sums the contents of ACC1 (from the previous cycle) with the output of the multiplier, i.e., $x_0(0,1) \times w_0(0,2)$ and the result is stored in the corresponding SRAM address. Similarly, the contents of ACC0 (from the previous cycle) are added to the output of the second adder, containing $x_0(0,1) \times w_0(0,1)$, and the result is stored in ACC1. In this cycle, MUX A0 passes the value 0 and consequently, the result of the first multiplier, $x_0(0,1) \times w_0(0,0)$ is directly stored in ACC0.

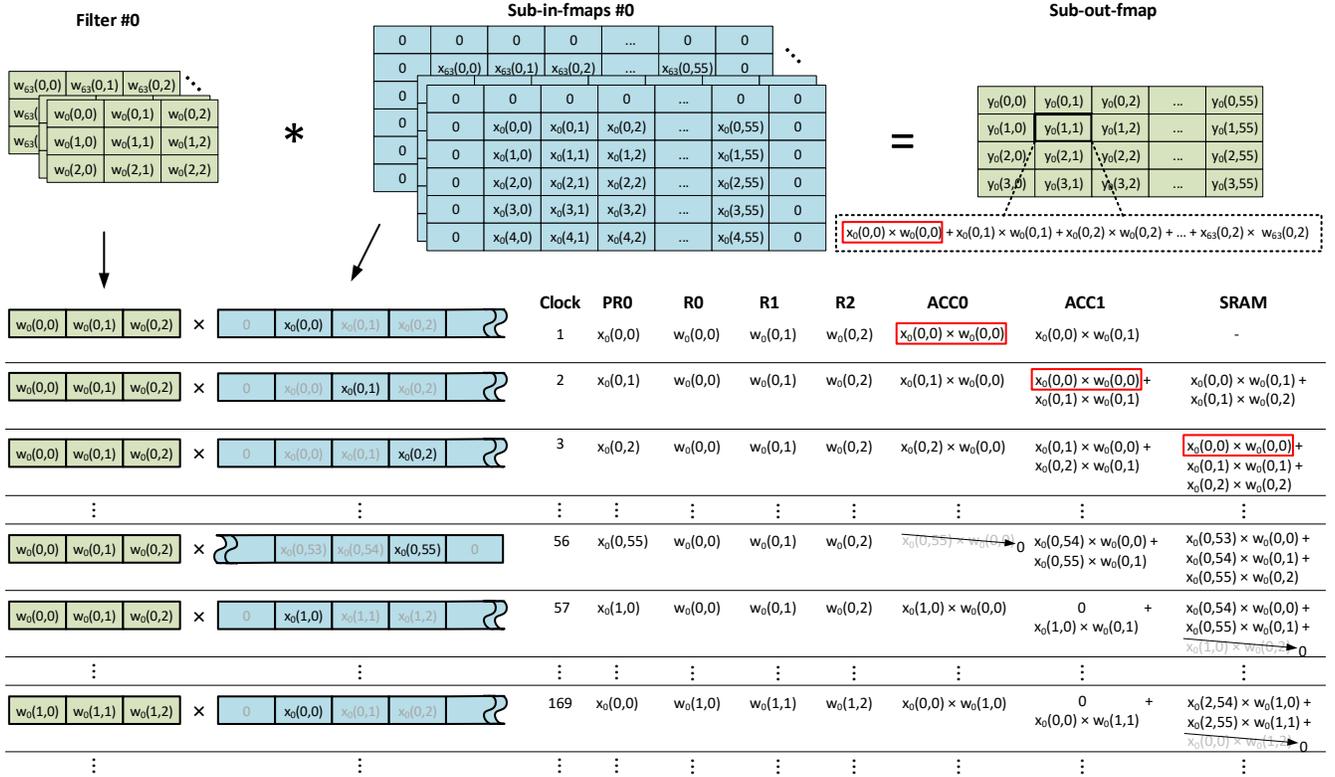

Fig. 5. The proposed dataflow for 3×3 convolution to generate a sub-out-fmap.

As shown with a red box in Fig. 5, the movement of $x_0(0,0) \times w_0(0,0)$ is completed in clock cycle #3 and the partial result of $x_0(0,0) \times w_0(0,0) + x_0(0,1) \times w_0(0,1) + x_0(0,2) \times w_0(0,2)$, is stored in SRAM. This is exactly the convolution of the first filter row with the first three elements of the sub-in-fmaps. This computation flow continues, as one input feature is fetched in each clock cycle, its element-wise multiplication with the weights of one filter row is performed, and the generated partial sums move across accumulator registers to finally be stored in the on-chip SRAM. As a result, all PEs actively contribute in computations in all clock cycles.

**In-fmap boundary handling:** In clock cycle #56, the input reaches the end of the first line, i.e., $x_0(0,55)$. Since there is no overlap between $x_0(0,55)$ and $w_0(0,0)$ their product does not need to be computed. Therefore, MUX M0 replaces this output with zero. The next input row starts with zero due to zero-padding. The produced zero in PE #0 plays the role of that zero-pad element of the next row in subsequent cycles. Without this mechanism, the architecture would have to fetch an additional zero as input, which would waste one clock cycle.

In clock cycle #57, $x_0(1,0)$ from the second row of the sub-in-fmaps, is fetched to start convolution with the first filter row. Since the product of $x_0(1,0)$ and $w_0(0,2)$ is not useful, MUX M2 substitutes the third multiplier output with zero. This zero plays the role of the zero-pad element of the last row, which saves another clock cycle. This zero-pad substitution mechanism, applied in clock cycles #56 and #57, save two clock cycles each time the computation reaches the end of an input row.

**Row filter and input channel change:** In clock cycle #169, the convolution of the first filter row with the first portion of sub-in-fmaps has been completed. Thus, the weights of the second filter row are stored in R0 to R2 registers and its convolution with the corresponding portion is started in a process similar to the first filter row. Without any stall, the computations are continued for other filter rows and along the input channels. The convolution in CU #0 is completed in clock cycle #39424, by fetching the last element of the sub-in-fmaps, $x_{63}(4,55)$, and storing the generated partial results in the on-chip SRAM in the next clock cycle.

*2) 3×3 Convolution Architecture Performance Analysis:* The computation time and the number of DRAM accesses for fetching input features and filter weights can be analyzed for the 3×3 convolution case. These deterministic metrics depend on the architecture and CNN characteristics such as number of CUs, number of filters, filter size, feature map size and number of input channels.

In CARLA, in each clock cycle, an input feature is read from the off-chip DRAM and a partial result is generated. Considering three filter rows and *IC* input channels, generating one sub-out-fmap in Fig. 5 requires $\frac{OL^2}{P} \times 3 \times IC$ clock cycles. This process is repeated for the $P$ partitions to generate the entire out-fmap. No clock cycle is spent for zero pad rows, which saves $2Z \times OL$ clock cycles. Moreover, the proposed boundary mechanism introduces no timing overhead for the zero pad columns.

CARLA uses $U$ parallel CUs, each computing one output channel in parallel. Thus, for $K$ filters (i.e., $K$ output channels), the computations are repeated $\left\lceil \frac{K}{U} \right\rceil$ times. Hence, the total number of clock cycles required to generate the entire out-fmaps is

$$\#Clock\ Cycles = (3 \times OL^2 - 2Z \times OL) \times IC \times \left\lceil \frac{K}{U} \right\rceil \quad (2)$$

The total number of DRAM accesses is composed of the accesses required to fetch the elements of in-fmaps and the filter weights and to store the out-fmaps. In 3×3 convolution, three input rows are required to generate an output row. Therefore, sub-in-fmaps have two more rows, i.e., (IL/P + 2), compared to sub-out-fmpas. For the $P$ partitions, the number of memory accesses for fetching input features, $\#DRAM\_Access_{in-fmaps}$, is obtained as

$$\#DRAM\_Access_{in-fmaps} = (IL + 2P - 2Z) \times IL \times IC \times \left\lceil \frac{K}{U} \right\rceil \quad (3)$$

The number of DRAM accesses required to store the results, $\#DRAM\_Access_{out-fmaps}$, is equal to the size of the out-fmaps, i.e., $OL^2 \times OC$. On the other hand, the architecture fetches three weights of a filter row in each cycle. This process is repeated for all $Q$ steps required to compute one sub-out-fmap. Having $U$ convolution units, the number of DRAM accesses for fetching weights of $K$ filters is increased by $\left\lceil \frac{K}{U} \right\rceil$ times. This process is repeated for the other partitions $P$ times. Hence, the total number of required DRAM access for fetching weights is

$$\#DRAM\_Access_{filter} = 3 \times U \times Q \times \left\lceil \frac{K}{U} \right\rceil \times P \quad (4)$$

One important metric in CNN hardware design is PE Utilization Factor (PUF), which indicates the proportion of time that PEs are actively contributing in computations [32].

$$PUF(\%) = \frac{\#Operations}{\#PEs \times \#Clock\ cycles} \times 100 \quad (5)$$

The number of operations in each convolutional layer is the total number of MAC operations required to compute out-fmaps excluding the operations associated to zero-pads. Thus, the total number of operations is

$$\#Operations = IC \times K \times (FL^2 \times OL^2 - 2Z \times (2 \times FL \times OL - 2Z)) \quad (6)$$

Replacing the number of clock cycles with (2) and the number of operations with (6), the PUF in (5) is obtained as $\frac{K}{(U+1) \times \left\lceil \frac{K}{U} \right\rceil}$. In the proposed design, the $U$ is set to 64 and $K$ is divisible to $U$. Thus, the PUF of the 3×3 convolution is 98.46%.

### B. 1×1 Convolution Mode

The proposed architecture could be used naively to perform 1×1 convolution by applying a dataflow similar to the one used in Section III.A. However, this would result in a considerable degradation of the PUF. Each input feature would be convolved with only one weight instead of three per row, and only one PE in each CU would be usable. The other two PEs per CU would be permanently inactive. To avoid this, CARLA exploits the parallelism available in the 1×1 convolution case using a distinct dataflow and hardware configuration.

In 1×1 convolution, the filter row size shrinks to 1. Thus, the computed partial sum in each PE does not have to be summed with the adjacent PEs. In other words, in 1×1 convolution, all PEs in CUs operate independently. On the other hand, in many convolutional layers, the number of features in each input channel is larger than the number of PEs. This provides the following possibility to parallelize the computations.

Each filter weight is convolved with all input features of the same channel in 1×1 convolution. A solution to improve the efficiency is to fill the PE registers with input features and slide the filter weights over them (which is a reverse from the 3×3 convolution mode). With this change, the PE registers R0, R1 and R2 can each store one input feature. The weights pass through the pipelined registers shown in Fig. 2 and appear at the PR$n$ input of each PEs. In every cycle, each PE participates in computations by multiplying the available weight at its input and its registered feature and accumulates the result.

*1) Description of CU Functionality Using an Example:* In contrast to the 3×3 convolution mode where each CU computes one sub-out-fmap, in 1×1 convolution, each CU computes a portion of sub-out-fmap of all output channels. For 1×1 convolution, each sub-in-fmap in Fig.1 is divided into small groups each containing $N$ features. $K$ different filters slide over each group of input features and generate $K$ output channels each containing $N$ output features. For instance, for $K = 64$ filters and a group size of $N = 3$ features, totally 3×64 partial results are computed in each CU.

We use an example from the ResNet models to describe the functionality of the CUs when computing 1×1 convolutions. In this example, the input feature map is of size 56×56×256, with no zero padding. There are 64 filters of size 1×1×256, the filter stride is 1, and the output size is 56×56×64. The architecture has 65 CUs (U+1=65) that together provide 196 PEs, because the first 64 CUs contain three PEs while the last CU has four PEs. The out-fmaps are partitioned into 56×56/196=16 sub-out-fmaps, each containing 196 features. In 1×1 convolutions, the number of features in sub-in-fmaps is the same as sub-out-fmaps, i.e. 196, and they are placed in the registers inside the 65 CUs while the filter weights are fed through the pipeline.

Fig. 6 shows the proposed dataflow to perform 1×1 convolution in CU #0. There are two types of working clock cycles in the architecture: routine computing cycles and exceptional cycles for the last input feature.

**Routine computing cycles:** In each routine computing cycle, three input features and one filter weight are read from the off-chip DRAM. The input features are loaded to three registers of a CU and the weight is given to the pipeline input. In clock cycle #1 in CU #0, the input PR0 is fed by the first weight of the first filter $w_0^0(0,0)$. This weight is multiplied by three input features, $x_0(0,0)$ and $x_0(0,1)$ and $x_0(0,2)$, which have been already loaded to the registers. The results are stored in the SRAM-S0, SRAM-S1 and SRAM-S2. In clock cycle #2, the first weight of the second filter, $w_0^1(0,0)$, is given to PR0, while the first input features are still in the PE registers. The dot product results are stored in different addresses in SRAMs. In the following clock cycles, the weights of the next filters enter the pipelined registers while new input features are stored in the PE registers of next CUs.

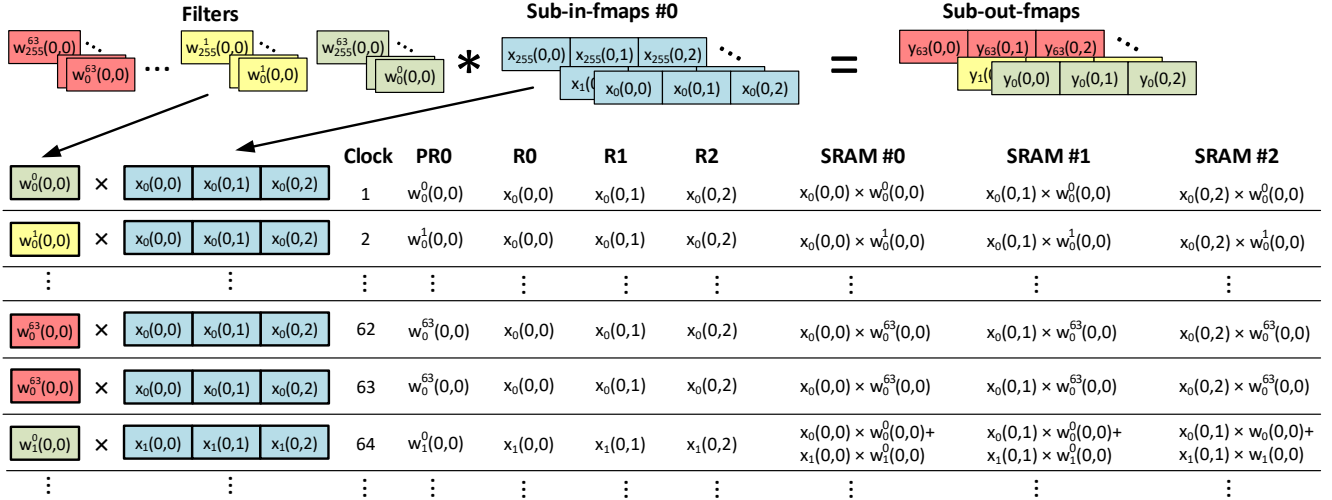

Fig. 6. The proposed dataflow for 1×1 convolution.

**Exceptional cycles for the last input feature:** The routine computing cycles continue until loading the input features for the last CU, #64, which exceptionally contains four PEs. To fill the registers of all these four PEs in one cycle, the architecture uses all four DRAM read buses to fetch four input features rather than three. Consequently, since no DRAM read bus is left to fetch a weight, the pipeline must be stalled in this cycle.

After this clock cycle, the dataflow returns to its normal operation cycle by fetching three features and one weight from the next input channel. The pipeline must be stalled each time the input feature fetch reaches the last CU. After $IC$ steps, each one taking many clock cycles, a sub-out-fmaps is generated and transferred to the off-chip DRAM.

*2) 1×1 Convolution Architecture Performance Analysis:* In the proposed dataflow, $U+1$ clock cycles are spent to perform the operations in an input channel and $IC$ steps are required to complete computation of one sub-out-map. Since there are $P$ partitions, this process is performed $P$ times to generate all entire out-fmaps. For $K$ filters, the computations are repeated $\lceil \frac{K}{U} \rceil$ times. Thus, the total number of clock cycles for 1×1 convolution is

$$\#Clock\ Cycles = (U+1) \times IC \times P \times \left\lceil \frac{K}{U} \right\rceil \quad (7)$$

One filter weight is fetched from DRAM every clock cycle, excluding the exceptional cycles. Thus, the total number of memory accesses for fetching filter weights is

$$\#DRAM\_Access_{filter} = U \times IC \times P \times \left\lceil \frac{K}{U} \right\rceil \quad (8)$$

The architecture must also fetch $\frac{OL^2}{P}$ features of a sub-in-fmaps and store them in CU registers. As this process is repeated for all $P$ sub-in-fmaps, the total number of DRAM accesses for fetching input features is

$$\#DRAM_{Access_{in-fmaps}} = OL^2 \times IC \times \left\lceil \frac{K}{U} \right\rceil \quad (9)$$

In CARLA, the pipeline must be stalled in exceptional cycles to fetch the last input features. In the example design, one stall cycle occurs in each 65 clock cycles. The PUF in this mode is obtained as $\frac{U}{U+1}$. For $U = 64$, the PUF is 98.46%.

*C. 1×1 Convolution Mode for Very Small in-fmap Size*

The proposed dataflow in Section III.B can perform 1×1 convolutions efficiently and achieves a high PUF when the number of features in each in-fmap is close to or greater than the number of PEs. If the number of features in a channel is radically lower than the total number of PEs, however, its effectiveness is degraded. For instance, in the last convolutional layers of the ResNet models, the size of the input feature maps shrinks to 7×7 with only 49 features in each input channel. In this case, only 49 PEs out of 196 would be utilized, resulting in a maximum PUF of 25%.

To improve the PUF, the architecture employs a distinct dataflow in which the weights reside inside CU registers and input features are fed into the pipeline. Using this approach, one limiting factor is that the one filter does not contain enough weights to fill all the PE registers in a CU. To overcome this limitation, we fill the remaining PEs with weights from other filters. This strategy is applicable due to the two following facts. (1) As in 1×1 convolution, PEs inside each CU operate independently, the registers in a CU can store weights from different filters. (2) In state-of-the art CNNs, the size of the in-fmaps typically shrinks by advancing in layers, while the number of filters increases. In this case the total number of clock cycles is

$$\#Clock\ Cycles = U \times IC \times \left\lceil \frac{K}{3U} \right\rceil \quad (10)$$

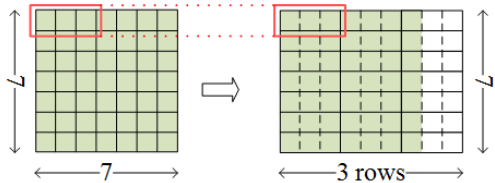

Fig. 7 Splitting each plane of a 7×7 filter into 21 pieces.

In this dataflow, each filter weight is only fetched once. Thus, the number of required memory accesses for fetching filter weights is

$$\#DRAM\_Access_{filter} = K \times FL^2 \times IC \quad (11)$$

The input features, however, need to be fetched multiple times if the number of weights is more than the number of PEs. Thus, the number of DRAM accesses to fetch input feature is

$$\#DRAM\_Access_{in-fmaps} = IL^2 \times IC \times \left\lceil \frac{K}{3U} \right\rceil \quad (12)$$

*D. 7×7 Convolution Mode and Others*

The architecture of CARLA is designed to achieve high PUF for 3×3 and 1×1 convolutions, as the most widely-used convolutions in most modern CNNs. However, other convolution sizes are used in exceptional cases. For example, the first layer in ResNet performs 7×7 convolutions. To support 7×7 and other larger convolutions, we divide the filter plane into smaller pieces and utilize a row-wise dataflow.

As shown in Fig. 7, the 7×7 filter is split into 21 pieces: 14 of them contain a row with three weights, and the remaining 7 consist of a row with one weight and two empty locations. The convolution of these small pieces can be handled by CARLA based on the row-wise convolutions explained in Section III.B. To preserve computation flow homogeneity, all 21 pieces with one weight are computed using the 3×3 convolution mode.

## IV. RESULTS AND DISCUSSION

CARLA was modeled in Verilog HDL and was implemented in TSMC 65 nm LP CMOS technology for a 200 MHz clock frequency using Cadence Genus. The word lengths of the weights, in-fmaps and out-fmaps were set to 16 bits while the accumulators were 24 bits wide. CARLA was evaluated with the convolutional layers of the original ResNet-50 and a structured sparse ResNet-50 model with 50% channel pruning rate.

*A. Specification of ResNet-50 and its structured sparse model*

ResNet has been the default CNN choice for image recognition since 2016 [14]. Table I illustrates the structure of ResNet-50 as one of the widely used ResNet models for benchmarking. In this model, the convolutional layers are grouped into five categories, Conv1 to Conv5, based on the size of layer outputs. All categories except Conv1, which only contains one 7×7 convolutional layer, are repeated several times. 32 out of the 49 convolutional layers require 1×1 convolutions while 16 layers need 3×3 convolutions.

Table I also shows the number of filters in ResNet-50 after applying a structured filter pruning method. Structured filter pruning is one of the promising solutions to reduce CNN model complexity that does not require indexing and keeps CNN model structure unchanged while reducing the number of filters. Since removing each filter leads to removal of one output channel, this approach is also called channel pruning [36]. We evaluated the number of DRAM accesses and computation time in CARLA with the structured sparse ResNet-50 model from a state-of-the-art filter pruning method [36]. The sparse model has 50% fewer channels while its accuracy only drops by 0.6 % compared to the original ResNet-50 model.

TABLE I
STRUCTURE OF RESNET-50 CONVOLUTIONAL LAYERS

| Convolutional layers | Output size | Filter size | # Filters in original model | # Filters in sparse model |
|---|---|---|---|---|
| Conv1 | 112×112 | 7×7 | 64 | 64 |
| 3 × Conv2 | 56×56 | 1×1 | 64 | 32 |
|  |  | 3×3 | 64 | 32 |
|  |  | 1×1 | 256 | 256 |
| 4 × Conv3 | 28×28 | 1×1 | 128 | 64 |
|  |  | 3×3 | 128 | 64 |
|  |  | 1×1 | 512 | 512 |
| 6 × Conv4 | 14×14 | 1×1 | 256 | 128 |
|  |  | 3×3 | 256 | 128 |
|  |  | 1×1 | 1024 | 1024 |
| 3 × Conv5 | 7×7 | 1×1 | 512 | 256 |
|  |  | 3×3 | 512 | 256 |
|  |  | 1×1 | 2048 | 2048 |

*B. Performance and DRAM Access Results*

Fig. 8 to Fig. 10 show the results of evaluating different efficiency metrics for CARLA, including PUF, computation time and DRAM memory accesses.

As shown in Fig. 8, CARLA achieves a remarkable PUF of 98% for 3×3 convolutions in all the convolutional layers of ResNet-50. For 1×1 convolution, the PUF is 98% for the Conv1 to Conv4 layers. In the Conv5 layer, in which the size of in-fmap is very small, CARLA utilizes the dataflow described in Section III.C and reaches a remarkable PUFs of 87.1% and 94.5%. The high PUFs for 1×1 and 3×3 convolutions highlight the effectiveness of the proposed dataflows to map the required computations on the available resources in most convolutional layers. The PUF for Conv1 with a single 7×7 convolution layer is only 45%. However, this lower PUF does not significantly impact the overall efficiency because it only applies to the first layer of ResNet-50. On the other hand, computing the 7×7 convolution using the method described in Section III.D, rather than using an additional customized architecture, prevents considerable hardware overhead.

Fig. 9 reports the computation time of CARLA for different convolutional layers of ResNet-50. The results show that 3×3 convolutions require more computations than 1×1 convolutions. The computation times for layers in Conv2 to Conv5 follow the same pattern. Inside each group of layers, the number of filters is inversely proportional to the number of input channels from one convolutional layer to the next. These two factors compensate each other with respect to the computation time. On the other hand, when moving from one group to the next, the out-fmaps size decreases while the number of filters increases proportionally. Consequently, and as the results in Fig. 9 illustrate, CARLA obtains similar computation times for the same filter size in various layers. To reduce the out-fmap size, the ResNet model utilizes filters with a stride of 2 between each convolutional group from Conv2 to Conv5. The computation times for those transition layers, i.e., convolutional layers #11, #23, and #41, are half of the values shown in Fig. 9 for the beginning layers of each group.

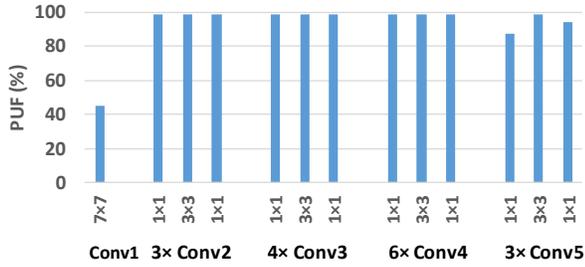

Fig. 8 The PUF for convolutional layers in ResNet-50.

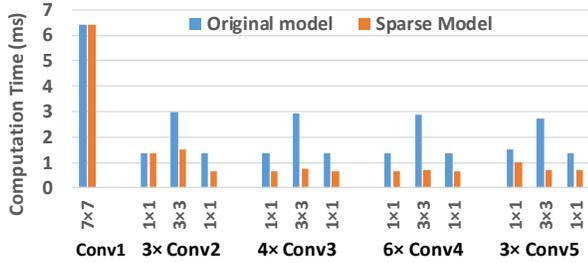

Fig. 9 The computation time for convolutional layers in ResNet-50.

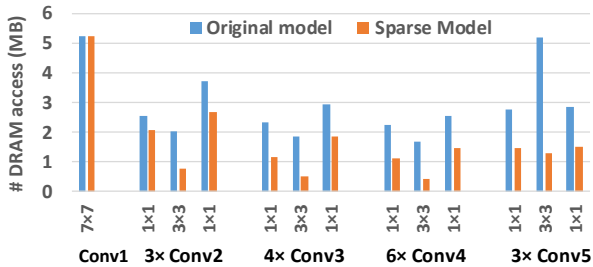

Fig. 10 # DRAM accesses for convolutional layers in ResNet-50.

Fig. 9 also shows the processing time reduction in CARLA for each convolutional layer of the structured sparse ResNet-50. In almost all convolutional layers, when the number of filters is reduced to half using pruning, CARLA achieves 2× to 4× speedup compared to the original ResNet-50 model. As mentioned earlier, pruning one filter results in removing one output channel, i.e., one input channel for the next layer. Therefore, in the layers without pruned filters, CARLA achieves a 2× speedup because of the reduced number of input channels. When both the number of input channels and the number of filters is halved, CARLA achieves 4× speedup. Overall, CARLA can perform the convolution computations of the original ResNet-50 in 92.7 ms and of the structured sparse ResNet-50 in 42.5 ms.

Fig. 10 shows the total number of DRAM accesses, which is mainly greater in Conv5 layers than in other layers due to their higher number of filters. In addition, the number of DRAM accesses for the Conv2 layers is larger due to its larger output size. For the Conv4 layers, where the output size is 14×14, the number of DRAM accesses is lower than for other convolution groups. This is a consequence of proper SRAM size selection. Since Conv4 is the most often repeated group of layers in ResNet-50 (6 times), we set the SRAM size to 224, which is divisible by all row sizes in ResNet-50, and greater than the number of generated partial results in the Conv4 layers, i.e. 14×14=196. This allows to maintain all partial results in on-chip memories. For other layers, due to the large size of the outputs, partitioning is unavoidable.

Fig. 10 also shows the number of DRAM accesses when running the convolutional layers of the sparse ResNet-50 by CARLA. As the results show, the number of DRAM accesses in CARLA continuously decreases in the convolutional layers of the structured sparse model. This is achieved by avoiding fetching pruned filters and by avoiding input feature re-fetches for the pruned computations. In addition, the number of generated output channels in the sparse model is lower and it requires fewer DRAM accesses. In other words, pruning a set of filters reduces the number of DRAM accesses required for filter weights, input and output features. Therefore, savings in the total number of DRAM accesses in CARLA are greater than the obtained reduction by pruning filter weights. The results in Fig. 10 also show that the reduction in the number of DRAM access is more noticeable in the Conv5 layers since most of the DRAM accesses in these layers are filter weight fetches. Overall, CARLA requires 124 MB DRAM accesses to perform the computations of the original ResNet-50 model and 63.3 MB DRAM accesses for the structured sparse ResNet-50. The discussion in this section can be generalized to the ResNet-101 and ResNet-152 models which have the same groups of convolutions with more iterations in each group.

### C. Comparison with State-of-the-Art CNN Implementations

Several existing works on hardware accelerator design for CNNs have targeted VGGNet-16. While CARLA supports all the variations across convolutional layers of ResNet-50 as a widely used CNN model, we also provide the evaluation results on VGGNet-16 to facilitate comparison with existing designs. Table II compares the implementation results of CARLA with the state-of-the-art. Two implementation metrics are of profound importance when comparing low-energy accelerators. First, the PUF measures dataflow efficiency for mapping the computations onto available PEs. Second, the number of DRAM accesses is critical and must be minimized since DRAMs consume orders of magnitude more energy than other functions.

Eyeriss [25] utilizes an NOC-like structure to compute VGGNet-16 and processes batches of three images at a time to reduce DRAM accesses. Data batching is a suitable technique to train neural networks. When computing inference, however, many real-world applications demand to process each image frame individually to make a quick decision instead of waiting for batch processing. The excessive latency of batch processing is a prohibitive factor for many real-time applications [15]. Eyeriss suffers from a high latency of 4.3 s for processing VGGNet-16 and obtains a PUF of 26% for VGGNet-16. CARLA achieves 11× faster image recognition with 72% higher PUF in 3×3 convolutions while consuming 31% smaller silicon area. Moreover, CARLA offers 3.6× higher performance and 3.5× higher energy efficiency thanks to its highly optimized model-specific dataflows.

TABLE II
COMPARISON OF THE PROPOSED METHOD WITH THE STATE-OF-THE-ART

|  | Eyeriss[25] | Envision [24] | FID [26] | ZASCAD [28] |  | CARLA (This work) |  |
| --- | --- | --- | --- | --- | --- | --- | --- |
| Technology (nm) | 65 | 28 | 65 | 65 |  | 65 |  |
| On-chip SRAM (KB) | 181.5 | 86 | 86 | 36.9 |  | 85.5 |  |
| Core Area (mm$^2$) | 12.25 | 1.87 | 3.5 | 6 |  | 6.2 |  |
| Gate Count (NAND2) | 1852 k | 1950 k | 1117 k | 1036 k |  | 938 k |  |
| Frequency (MHz) | 200 | 200 | 200 | 200 |  | 200 |  |
| Batch size | 3 | N/A | 1 | 1 |  | 1 |  |
| Bit precision | 16b | 1-16b | 16b | 16b |  | 16b |  |
| #PEs | 168 | 256-1024 | 192 | 192 |  | 196 |  |
| CNN model | VGG-16 | VGG-16 | VGG-16 | VGG-16 | ResNet-50 | VGG-16 | ResNet-50 |
| Power (mW) | 236 | 26 | 260 | 301 | 248 | 247 | 247 |
| PUF % (filter size) | 3×3: 26% | 3×3: 32% | 3×3: 89 % | 3×3: 94% | total: 88% | 3×3: 98% | 1×1, 3×3: 98% 7×7: 45% |
| Latency (ms) | 4309.5 | 598.8 | 453.3 | 421.8 | 103.6 | 396.9 | 92.7 |
| Performance (Gops) | 21.4 | 51.3 | 67.7 | 72.5 | 74.5 | 77.4 | 75.4 (83.26)† |
| Efficiency (Gops/W) | 90.7 | 1973 | 260.4 | 240.9 | 300.4 | 313.4 | 305.3 (337.1)† |
| # DRAM access/batch (MB) | 321.1 | N/A | 331.7 | 375.5 | 154.6 | 258.2 | 124.0 |

† In [28], the number of operations (Gop) in ResNet-50 is different of our calculations. The number in parenthesis are computed when using the same number of operations as [28].

Envision [24] takes advantage of the more advanced 28 nm UTBB FD-SOI technology and dynamic voltage-accuracy frequency scaling (DVAFS) to improve its energy efficiency. That design offers a relatively low PUF of 32% resulting in a latency of 598.8 ms when running VGGNet-16. CARLA outperforms the Envision architecture by achieving 34% lower latency while requiring 2.1× fewer gates. The number of DRAM accesses has not been reported for Envision.

FID [26] uses a dataflow inspired from the computational pattern of FC layers to compute the convolutional layers of VGGNet-16. Fig. 11 compares the number of DRAM accesses in CARLA with FID for each convolutional layer of VGGNet-16. As these results show, our proposed pipelining scheme significantly improves data reuse in most convolutional layers and consequently reduces the number of re-fetched input features. A large portion of the total DRAM accesses is associated with fetching the input features. Hence, in CARLA, the achieved reduction in the number of DRAM accesses for fetching input features reduces the total number of DRAM accesses per layer. Overall, compared to FID, CARLA reduces the latency by 12.4% and the number of DRAM accesses by 22.1%.

ZASCA [28] is built upon FID to support a larger number of filter sizes. Compared to FID, ZASCA requires 2× larger silicon area and can perform the convolutional layers of ResNet-50. The ZASCA configuration for computing the original CNN model (dense activations) is called ZASCAD in [28]. Fig. 12 and Fig. 13 compare the PUF obtained by ZASCAD and CARLA when computing 3×3 and 1×1 convolutional layers, respectively.

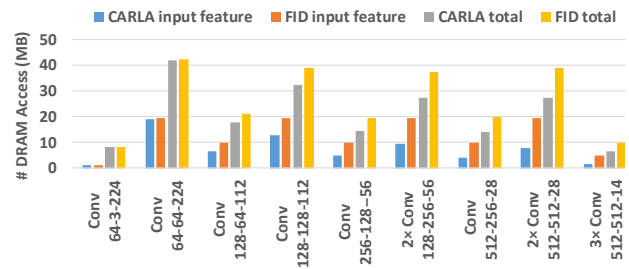

Fig. 11 The number of DRAM accesses for CARLA compared to FID for VGGNet-16. Conv 64-3-224 means a convolutional layer with 64 filters, 3 input channels, and in-fmap size of 224.

As shown in Fig. 12, our proposed dataflow for 3×3 convolutions outperform the FC-inspired dataflow in ZASCAD especially in early convolutional layers. The FC-inspired dataflow in ZASCAD suffers from an issue called weight passing [27], which consists of wasted clock cycles to correct weight positions in the weight generator unit when computing a new output feature row. In addition, in ZASCAD there is no mechanism to enable performing convolution and DRAM data transfers simultaneously. The lack of such a mechanism introduces interrupts in the computations in ZASCAD, reducing the PUF and increasing the computation time.

In CARLA, the proposed dataflow for 3×3 convolutions can utilize all the processing elements in all computation cycles and takes advantage of a mechanism to handle zero paddings at boundaries. Thus, it does not waste any clock cycle. In addition, utilizing paired SRAMs enables CARLA to overlap computations with SRAM-DRAM transfers. This results in a near perfect PUF of 98%.

As shown in Fig. 13, the PUF of ZASCAD for 1×1 convolution is severely degraded in the majority of the convolutional layers. The 1×1 dataflow in ZASCAD does not efficiently map the computations to available PEs and many of them are idle. For instance, in the convolutional layer #2 (L2), 128 out of 192 of PEs do not contribute in computations [27].

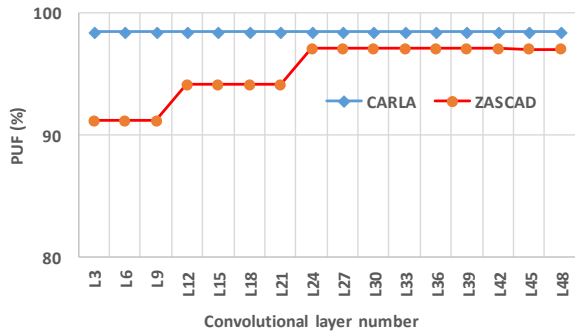

Fig. 12 The PUF for CARLA compared to ZASCAD for 3×3 convolutional layers in ResNet-50. L3 means the convolutional layer #3. The ZASCAD architecture was previously introduced under the name MMIE in [27]. Data for ZASCAD (MMIE) are collected from [27].

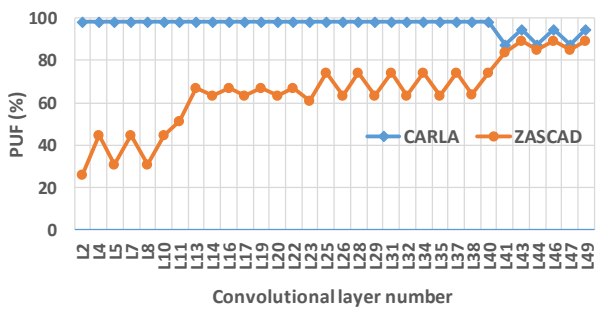

Fig. 13 The PUF for CARLA compared to ZASCAD (MMIE) for 1×1 convolutional layers in ResNet-50.

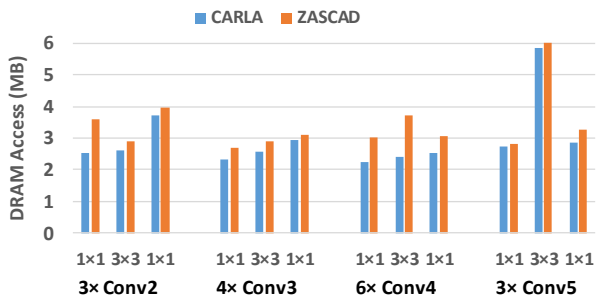

Fig. 14 The number of DRAM accesses for CARLA compared to ZASCAD (MMIE) for ResNet-50.

The reconfigurable architecture of CARLA, on the other hand, maintains computation efficiency high by using a different dataflow for 1×1 convolution. As discussed in sections III.B, PEs have valid data to perform computations and they remain active in most computational cycles of the 1×1 convolution mode.

Fig. 14 shows the number of DRAM accesses for 1×1 and 3×3 convolutions in CARLA and ZASCAD. CARLA requires fewer DRAM accesses for both 3×3 and 1×1 convolutions. In 1×1 convolution, the lower PUF of ZASCAD causes a larger number of re-fetches of the same data from the off-chip DRAM. In 3×3 convolution, similar to the discussion for VGGNet-16 in Fig. 11, CARLA requires fewer DRAM accesses thanks to the employed feedback mechanism in its pipeline. Overall, CARLA requires 19.8% fewer DRAM accesses compared to ZASCAD, while offering 10.5% lower latency for computing convolutional layers of ResNet-50.

When comparing CARLA on VGGNet-16 benchmark with [24]-[26] in Table II, we assume 64-bit data bus width for off-chip DRAM in conformance to their assumptions. However, our evaluation show that the implementation results for ZASCAD in [28] on ResNet-50 benchmark can be obtained only by assuming a wider data bus for DRAM. It is due to characteristics of Conv5 layers in ResNet-50. In this group of convolutional layers, the number of input features in a channel is 7×7= 49 which is radically lower than the total number of PEs, i.e., 192. In each fetch from DRAM, in [28], one 16-bit input feature along with three filter weights are read. After fetching all the input features within 49 clock cycles, there are still plenty of PEs which are empty and not loaded with filter weights. On the other hand, other PEs which were previously loaded by weights are data-starved since all the corresponding input features have been already fetched. To make all the PEs active and achieve a high PUF, the accelerator has to fill the empty PEs with three filter weights and simultaneously fetch a new set of three filter weights and one input feature for the data-starved PEs. Hence, assuming 16-bit word-length for data values, in total $3×16 + 4×16 = 112$ bits must be fetched from DRAM in each clock cycle to achieve a high PUF. We used this assumption when comparing CARLA with [28] on the ResNet-50 benchmark. If the DRAM bus width shrinks to the tight assumption of 64-bit, CARLA will need some stalls to first feed all CUs with input features and then starting the computation of the next filter row. The computation time in that case is 98.2 ms, which is still faster than ZASCAD. This highlights the advantage offered by dataflows even when dealing with low bandwidth DRAMs. The DRAM data bus width does not impact the number of DRAM accesses and CARLA still requires 19.8% fewer DRAM accesses.

## V. CONCLUSION

This work introduced CARLA, which is a convolutional accelerator with several operating modes to efficiently compute the convolutions in very deep CNNs such as ResNet-50. For 3×3 convolution, CARLA utilizes a new dataflow to maximally reuse the data fetched from DRAM. In addition, it utilizes an effective solution to avoid latency overhead while switching the rows and channels in boundaries of input feature maps. To support 1×1 convolution, CARLA maximizes data reuse by switching the datapath of the input data and the filter weights. When evaluated in ResNet-50, this architecture achieves a PUF of 98% across many of 1×1 and 3×3 convolutions and supports other convolution structures such as 7×7. This results in a latency of 92.7 ms and total DRAM accesses of 124.0 MB when computing the convolutional layers of ResNet-50 for classifying an input image. In addition, CARLA reduces the latency to 42.5 ms and the required DRAM accesses to 63.3 MB when half of the channels are pruned.